\documentclass[pra,superscriptaddress,showpacs,twocolumn]{revtex4}

\usepackage{amsmath,amssymb} 
\usepackage{graphicx,color}
\usepackage{dcolumn}

 \newcommand{\ket}[1]{|{#1}\rangle}

 \newcommand\bracket[2]{\langle #1 | #2 \rangle}

 \newcommand {\be} {\begin{equation}}
\newcommand {\ee} {\end{equation}}

\newcommand{\U}{\hat{U}}
\newcommand{\C}{\hat{C}}

\newcommand{\Pa}{\hat{P}}

\newcommand{\Sa}{\hat{S}}

\begin{document}
    
\title{Randomized benchmarking of single and multi-qubit control in liquid-state NMR quantum information processing.}
\author{C.~A. Ryan\footnote{These authors contributed equally to this work.}}
\affiliation{Institute for Quantum Computing and Dept. of Physics, University of Waterloo, Waterloo, ON, N2L 3G1, Canada.}
\author{M. Laforest$^*$}
\affiliation{Institute for Quantum Computing and Dept. of Physics, University of Waterloo, Waterloo, ON, N2L 3G1, Canada.}
\author{R. Laflamme}
\affiliation{Institute for Quantum Computing and Dept. of Physics, University of Waterloo, Waterloo, ON, N2L 3G1, Canada.}
\affiliation{Perimeter Institute for Theoretical Physics, Waterloo, ON, N2J 2W9, Canada}

\begin{abstract}
Being able to quantify the level of coherent control in a proposed device implementing a quantum information processor (QIP) is an important task for both comparing different devices and assessing a device's prospects with regards to achieving fault-tolerant quantum control.  We implement in a liquid-state nuclear magnetic resonance QIP the randomized benchmarking protocol presented by Knill et al (PRA \textbf{77:} 012307 (2008)).  We report an error per randomized $\frac{\pi}{2}$ pulse of $1.3 \pm 0.1 \times 10^{-4}$ with a single qubit QIP and show an experimentally relevant error model where the randomized benchmarking gives a signature fidelity decay which is not possible to interpret as a single error per gate.  We explore and experimentally investigate multi-qubit extensions of this protocol and report an average error rate for one and two qubit gates of $4.7 \pm 0.3 \times 10^{-3}$  for a three qubit QIP.  We estimate that these error rates are still not decoherence limited and thus can be improved with modifications to the control hardware and software.   
\end{abstract}  


\maketitle

\section{Introduction}
Quantum information processing devices have the potential to revolutionize our understanding of computational complexity and solve certain problems exponentially faster than current classical algorithms.  In order to achieve these goals the ability to coherently control a large number of two level quantum systems (qubits) will have to be demonstrated.  An important issue in this research path is to be able to quantify the level of control demonstrated.   A clear, systematic and standardized algorithm is needed to be able to report the relevant level of control achieved in a given system.  Such a protocol would be useful in a number of ways: it should provide a fair and transparent way to compare different devices and technologies; it should provide a way to quantify engineering improvements to the same device and it should provide a rough measure of the device's prospects with regards to fault-tolerant computation \cite{Preskill:1998a}.   

Full characterization of any quantum process, and hence calculation of the fidelity of control, is possible through a procedure known as quantum process tomography (QPT)~\cite{Neilson:1997a}.  However, there are a number of caveats with this approach.  It is difficult to analyse and to reconstruct a completely positive map from the results when there are errors in the preparation and readout steps and/or there is noise in the measurments \cite{PhysRevA.67.042322}.  Indeed to quantify the error in a certain gate with QPT, readout and preparation pulses with a lower error level than the gate being measured are required.  QPT gives full characterization of a particular quantum gate in a particular setting.  Although this is useful information, it does not necessarily tell us how another gate will perform, or even how the same gate will perform as part of a larger computation.  Finally, full QPT requires an exponential number of experiments, making it experimentally prohibitive for QIP's larger than a few qubits.    

Ultimately, full knowledge of a quantum operation is often not needed to provide an answer to the above problems.  Randomization has been proposed as a useful technique in revealing a smaller number of relevant  coarse-grained parameters of the channel \cite{Emerson:2003a}.  By twirling a channel with random, Haar distributed, unitaries the channel is reduced to a depolarizing channel with a single parameter to describe the strength of the noise and thus the average gate fidelity.  This approach benefits from the concentration of measure in large Hilbert spaces whereby the average fidelity can be estimated with only a few experiments \cite{Emerson:2005a}. This technique can be generalized to a sequence of random unitaries and a fidelity decay is measured as function of increasing number of gates.  The rate of fidelity decay can then be measured and related to the average gate fidelity. 

Generating fully Haar-random unitaries for this protocol is inefficient as it requires an exponential number of continuous parameters and thus an exponential amount of elementary gates to describe and create and Haar-random unitary gate.  Fortunately, previous work has shown that the Clifford group is a unitary 2-design, meaning it  is sufficient to sample from the $n$ qubit Clifford group to depolarize a $n$ qubit channel and to estimate its average fidelity \cite{Divencenzo:2002a,Dankert:2006a}.  Efficient methods exist for generating random Clifford gates from elementary 1 and 2 qubit gates \cite{Divencenzo:2002a,PhysRevA.70.052328} and it is even possible to reduce the number of gates required by using pseudo-random Clifford gates from either a prescribed algorithm \cite{Dankert:2006a}, or simply multiplying together randomly chosen 1 and 2 qubit gates \cite{Harrow:2008a}.  Randomized benchmarking of single qubit Clifford group gates was formalized in a protocol presented by Knill et al. \cite{knill:012307}, where the fidelity decay under a sequence of random Clifford group operations is measured and the average gate fidelity can then be calculated.

Liquid state NMR offers a clean system with high fidelity control built on decades of engineering experience in NMR spectroscopy.  Utilizing this control, liquid-state NMR QIP's have established many demonstrations of quantum algorithms and simulations \cite{negrevergne:170501,Vandersypen:2001a} and are an ideal testbed for exploring ideas about quantum control for quantum information processing purposes \cite{ryan:012328}.  Here we present results of applying these randomized benchmarking protocols to both single and multiple qubit gates in a liquid state NMR QIP.  In these experiments we are able to quantify the control achieved by both standard pulse techniques on a single qubit and more advanced pulse shaping approaches from optimal control theory in the multi-qubit setting.  While our single qubit experiments followed Ref. \cite{knill:012307}, there are potentially many generalizations of the protocol to more than one qubit and we suggest two such protocols.  Finally, it is difficult to obtain analytical results in the case of benchmarking pulse dependent errors.  Indeed, we find and analyze an experimentally relevant error model where randomized benchmarking fails to reveal a single average error per gate.  This serves to highlight the difficulty in devising universal efficient benchmarking protocols.  

\section{Protocols}
The protocols are a form of a generalized motion reversal applied to efficient gate fidelity estimation \cite{Emerson:2005a}.  The basic steps are to apply sequences of random unitary gates and then measure the average fidelity decay as a function of the number of gates. With the assumption that the errors are independent of the gate performed and that the gates are chosen uniformly according to the invariant Haar measure, the series of random gates and averaging over different gate sequences will effectively depolarize the noise.  That is, the state after a self-inverting sequence of $n$ gates is given by:
\be
\label{emersonderive}
\begin{split}
\ket{\rho(n)}\rangle &=\int \left[\Pi_i d\U_i\right] \hat{\Lambda}\U_n\ldots\hat{\Lambda}\U_2\hat{\Lambda}\U_1\ket{\rho(0)}\rangle  \\
&=\hat{\Lambda}_{ave}^n\ket{\rho(0)},
\end{split}
\ee
where $\ket{\rho(i)}$ is density matrix $\rho(i)$ after the $i$'th operation represented as a vector in Liouville space, $\U_i=U_i^*\otimes U_i$ is the superoperator representation of the unitary gate $U_i$  and $\hat{\Lambda}$ is the noise superoperator \cite{havel:534}.  Under the averaging $\hat{\Lambda}_{ave}$ becomes a depolarizing noise \cite{Emerson:2005a}, 
\be
{\Lambda}_{ave}(\rho)=p_{\Lambda}\rho+(1-p_{\Lambda})\frac{\openone}{D},
\ee
where $D$ is the dimensionality of the system and the depolarizing parameter is related to the original noise operator by
\be
p_\Lambda = \frac{\textrm{Tr}(\hat{\Lambda})-1}{D^2-1}.
\ee

Therefore, we expect the average fidelity of the output state with respect to an arbitrary input state after $n$ gates to decay exponentially to a saturation level which depends on the dimension of the Hilbert space:
\be
\overline{F}_n = p^n_{\Lambda}\left[\textrm{Tr}(\rho(0)^2)-\frac{1}{D}\right]+\frac{1}{D},
\ee
where we have defined 
\be
F_i =  \frac{1}{D}\bracket{\rho(i)}{\rho(0)} = \frac{1}{D}\textrm{Tr}[\rho(i)^\dagger\rho(0)].
\ee

Measuring the decay of the average fidelity thus gives us a concrete information about the strength of the noise, without giving the details of the action of the noise.  From an error correction and fault-tolerance perspective, the schemes are usually developed regardless of the specifics of the action of the noise and the strength of the noise is the most relevant piece of information.  And from the strength of the noise the average gate fidelity can be calculated:
\be
\overline{F_g}\left(\Lambda\right) = p_\Lambda + \frac{1-p_\Lambda}{D}.
\ee 

Because the gate fidelity corresponds to a second order polynomial in the gate and its complex conjugate   (also known as a $(2,~2)$ polynomial), the average gate fidelity over the Haar measure can be evaluated using a unitary 2-design so that the continuous integral over the unitaries can be replaced by a sum over, for example, the finite Clifford group $\mathcal{C}$ \cite{Divencenzo:2002a,Dankert:2006a}, i.e.
 \be
  \int d\U\,\U^{-1}\hat{\Lambda}\U = \frac{1}{|\mathcal{C}|}\sum_{\C\in\mathcal{C}}\C^{-1}\hat{\Lambda}\C.
\ee
  
Then the sequence of random unitaries becomes a sequence of random Clifford group gates.  The use of Clifford gates for benchmarking has a number of justifications.  Clifford group operations are of paramount  importance in most fault tolerant constructions based on stabilizer codes.  The Clifford group operations are the main computational elements and universality is granted via state preparation of so-called ``magic states", e.g. states of the form $\cos{\frac{\pi}{8}}\ket{0}+\sin{\frac{\pi}{8}}\ket{1}$ \cite{bravyi:022316}.  The performance of many computational steps can be bootstrapped through the use of higher fidelity Clifford group operations, e.g. several noisy magic states can be purified with ideal Clifford gates to create one magic state with a lower error rate \cite{bravyi:022316}.  Morevover, the state's evolution under Clifford group operations can be efficiently tracked classically allowing an efficient construction of a recovery gate and/or prediction of the ideal final state \cite{PhysRevA.70.052328}.  

The protocols are designed to extract the average gate fidelity which under reasonable assumptions about the error model should be the computationally relevant quantity.  Algorithms using only Clifford operations, for example many  fault-tolerant constructions, can be Pauli randomized at every step (whether it occurs inherently as part of a teleportation \cite{Knill:042322} or is explicitly put in) and so the quantity measured by randomized benchmarking should be close to the error rate experienced in an algorithm.     It is certainly true that with many qubits the Hilbert space is large enough to hide a worst case fidelity of 0 while the average fidelity is very high.  And so, it is possible that some very large, highly correlated and specially designed error will be undetected by this benchmarking procedure.  However, this would seem to require a contrived unphysical error model.  Furthermore, for one and two qubit gates the Hilbert space is too small for the worst case and average fidelity to be significantly different.  Finally because it is too difficult to show fault-tolerance for an arbitrary distribution of errors,  proofs \cite{Aliferis:2006a} and simulations \cite{Knill:2005a} of fault-tolerance schemes rely on a stochastic distribution of error locations or a depolarizing error model respectively, for which the average fidelity should be the relevant quantity to measure.

\subsection{Single Qubit}
In the case of the single qubit benchmarking we followed exactly the implementation of Knill et al \cite{knill:012307}.  For depolarizing one qubit noise, the single qubit Clifford operations are isomorphic to the 48 operations parametrized as 
\be \label{PauliSimplectic}
\begin{split}
\mathcal{C} &\cong \mathcal{S}\mathcal{P} \\
&=e^{\pm i\frac{\pi}{4}Q}e^{\pm i\frac{\pi}{2}P} ,
\end{split}
\ee
where $Q\in\{X,Y,Z\},\,\,P\in\{\openone,X, Y, Z\} $, that is, a $\pi$ pulse (or Pauli operation) followed by a $\frac{\pi}{2}$ pulse (or a symplectic operation). The symplectic operations are deemed the  ``computationally relevant'' operations that advance the computation while the Pauli operations serve only to redefine the Pauli frame.

\begin{figure}[htbp]
\includegraphics[scale=0.4]{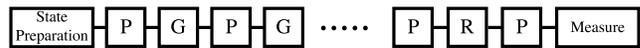}
\caption{\label{singlecircuit}
Quantum circuit implementing single qubit benchmarking.  A fiducial state is prepared and a sequence of computational gates $G$ is applied.  The recovery gate $R$ is chosen to return system in a known final state.  The Pauli gates $P$ interleaved with the computational gates induce a Pauli randomization.}
\end{figure}

The circuit implemented is shown in Figure \ref{singlecircuit}.  To perform an approximate averaging, a series of 192 computational gates was chosen at random and truncated at a series of different lengths.   Random Pauli operations were then inserted between each computational gate.  The initial state was chosen to be the thermal state in NMR:  $\frac{1}{2}\openone + \epsilon Z$ (where $\epsilon \approx 10^{-5}$).  The identity component is unobservable in NMR and can be considered a large error in the preparation or measurement which is normalized out by the protocol.  The state was tracked through the computational gates and the recovery gate R was chosen to return the state to either $+Z$ or $-Z$ with equal probability.  The state was then readout with a 90 degree readout pulse and the fidelity measured by comparing the integral of the signal to a reference spectrum. For each truncation, the Pauli operations were randomized 8 times.  Each point was further averaged over four different computational gate sequences and the averaged fidelity from the 32 experiments for each truncation was used in the fitting.

One technical point to note is that rotations about the $Z$ axis are implemented through an abstract frame change (changing the phase of subsequent pulses and potentially the observation) and take no time.  However, for consistency, a delay equivalent to the $\frac{\pi}{2}$ or $\pi$ pulse time was executed for those gates.  This is the procedure followed in \cite{knill:012307}.  However, performing the $Z$ rotation in this manner (as opposed to physically implementing the gate) is not as effective at depolarizing the noise because in commuting the $Z$ rotation through the pulse sequence it is also assumed the $Z$ rotation commutes with the noise operation.  In situations where the noise is dominated by dephasing this may be appropriate but for a general case this is not true.     

\subsection{Multiple Qubits}
In the case of more than one qubit, it is  difficult to prescribe the correct gate set for determining an error per gate.  The gates should depolarize the noise but at the same time the error per gate should be meaningful in relation to the fault-tolerant thresholds. It would be ideal to  quantify the error per gate for one and two qubit gates and also storage errors for wait steps.  However, it is difficult to isolate the errors for only these gates if the error model does not satisfy the independent error model - that each gates errors are described by a quantum operations acting only on qubits which the gate affects.  In realistic situations it is most likely that applying a gate to qubit $a$ could induce an error on qubit $b$.

One possibility is to choose a generating gate set consisting of single qubit Clifford generators (say the Hadamard and phase gates) and controlled NOT's between pairs of qubits.  This will generate the multi-qubit Clifford group and indeed after only a small number of gates will approximate a 2-design necessary for depolarizing the noise \cite{Harrow:2008a}.   The multi-qubit protocol then becomes:

\begin{enumerate}

\item Choose a series of lengths of computational gates to measure the fidelity decay at.  The number of random gates necessary to achieve depolarization of the noise depends on the number of qubits and may be large.  Thus we expect only the asymptotic error rate to be meaningful.

\item For each truncation length choose $n_g$ random sequences of computational gates from the generating set of the full $n$-qubit Clifford group.  

\item Determine a recovery sequence which will return the state to one with a known definite output upon measurement in the absence of error.  This can either undo the entire sequence to return to the input state or ensure that one stabilizer  has a certain measurement outcome as suggested in Ref. \cite{knill:012307}.    Because the Clifford group operations can be efficiently tracked this is possible to do efficiently on a classical computer and should have no more than $\mathcal{O}(n^2/\log n)$ gates \cite{PhysRevA.70.052328}. 

\item Apply some parallelization routine to the random sequence of gates to ensure that the number of wait steps does not grow with the size of the computer.  This parallelization step allows a fair comparison between different size QIP, say a 5 and a 50 qubit computer.  The error per time step may be larger in the 50 qubit computer but many more gates are possible in each timestep.  

\item Measure the fidelity decay as in the single qubit case.  An exponential fit to the fidelity decay will reveal the average error per one and two qubit gate.  It is possible that the average error could mask a distribution of error rates such that for example all single qubit gates are perfect but the two qubit gates are much worse.  However, more detailed, but still coarse grained information is available by doing more experiments (see Sec. \ref{multiqubit_extensions}).  

\end{enumerate}

Numerical simulations have confirmed that this protocol will return the correct asymptotic ( beyond $\approx 30$ gates for the 3 qubit case) error rate for a variety of error models such as dephasing and pulse dependent unitary errors. For the later, it should be mentioned that we made the assumption that the errors were of the same strength, hence numerically verifying the conjecture in Ref. \cite{Emerson:2005a} for this case.  Not surprisingly, larger amounts of randomization are required compared with the single qubit protocol.

\section{Experiment}
The experiments were performed in liquid state NMR on a 700MHz Bruker Avance spectrometer using a TCI cryogenic probe.  The cryo-probe provides enhanced sensitivity and associated  improved signal-to-noise ratio but the high quality factor of the probe resonant circuit leads to phase-transient and radiation damping effects.     

\subsection{Single Qubit} \label{single_exp}
The proton spins of unlabeled chloroform were chosen as the single qubits.  A sample was made from a 0.3\% aqueous solution of unlabeled chloroform dissolved in d6-acetone.  The sample was not vaccuum-pumped to avoid unnecessarily long $T_1$ relaxation times.  The $T_1$ was measured to be 7 seconds through inversion recovery and the $T_2$ to be 4.5 seconds using a standard CPMG sequence.  The unrefocussed $T_2^*$ was 0.45 seconds calculated from the spectral linewidth of the NMR signal.  
 
To address the amplitude and phase transient issues with the high Q cryoprobe, $24\mu s$ gaussian shaped $\frac{\pi}{2}$ pulses were used, which avoid these unwanted effects due to their more slowly varying amplitude profile. Since the largest part of the errors are expected to be due to pulse miscalibration, amplifier drift and r.f. inhomogeneity, composite pulses, robust to r.f. field variation were also tested.  The BB1 family of pulses from Wimperis et al. \cite{Wimperis:1994a} are robust to pulse length (calibration) errors $\epsilon$ up to order $\epsilon^6$ and are universally compensating in that they are robust unitary operations rather than robust for a particular state to state transformation.  Their usefulness in experimental QIP has been previously  reported \cite{xiao:032334}.  The pulses consist of a compensating block followed by the desired pulse so that a rotation by an angle $\theta$ about the $x$ axis can be replaced by,
\begin{equation}
R_x\left(\theta\right) = \left(180\right)_{\phi_1}\left(360\right)_{\phi_2}\left(180\right)_{\phi_1}R_x\left(\theta\right) .
\end{equation}

Where, $\phi_1$ and $\phi_2$ depend on the pulse flip angle: 
\begin{equation}
\phi_1 = \frac{1}{3}\phi_2 = \arccos\left(\frac{-\theta}{4\pi}\right).
\end{equation}

The location of the compensating block is not important and it can be placed before or after the pulse.  The pulse can even be symmeterized by placing the compensating block between two halves of the pulse \cite{PhysRevA.67.042308}.  

\begin{figure}
\includegraphics[scale = 0.35]{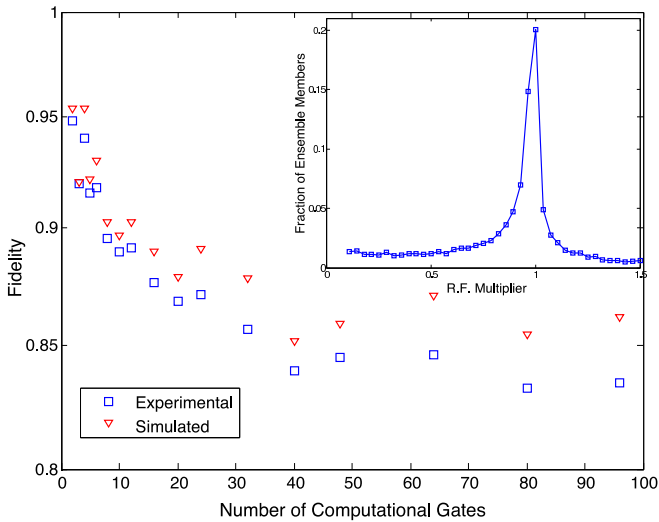}
\caption{\label{singlequbit_norf}
(Color online) Experimental ($\square$) fidelity as a function of number of randomized gates for a single qubit using BB1 composite pulses plotted on a semi-log plot.  The fidelity decay is clearly non-exponential indicating incoherent pulse dependent errors \protect{\cite{Henry:2007a}}.  This effect is caused by the large distribution of r.f. field strengths across the sample shown in the inset.  Also shown are the results from simulations of the pulse sequence ($\triangledown$) averaged over the measured r.f. profile.  The simulations match the experimental results both qualitatively and quantitatively.}
\end{figure}

The results of the single qubit benchmarking with BB1 composite pulses are shown in Figure \ref{singlequbit_norf}.  It is clear that the pulse fidelity is low and furthermore that the curve does not fit a single exponential decay well.  However, these results  can be explained by the r.f. field strength variation across the sample.  This r.f. inhomogeniety is particularly bad in cryogenic probes \cite{Kobzar:2004la}.   Indeed, by measuring the r.f. inhomogenity profile and simulating the experiment across that variation we were able to reproduce both quantitatively and qualitatively the results showing we understand well the error model.  The result can be interpreted intuitively in that we expect spins which see an r.f. field very different to the ideal field to very quickly end up at some random point on the Bloch sphere whereas those close to the ideal field strength will closely track the ideal evolution for many gates.  Thus we expect the fidelity to initially decay quickly (with large fluctuations) as the spins at the edge of the r.f. profile are depolarized and then for the fidelity to level off and decay much more slowly.  This intuitive picture is confirmed in a more detailed analysis in Appendix \ref{rfanalysis_appendix}.  It is also interesting to note that with this pulse-dependent coherent error model, it is impossible to average the fidelity decay to a single exponential; it is always a sum of exponentials with different decay rates.  This error model is not restricted to ensemble effects but would also apply in the case were a parameter (say a laser power in an ion trap) slowly varies so that it is constant for the time of one experiment but fluctuates from experiment to experiment. 

\begin{figure}[htb]
\includegraphics[scale = 0.35]{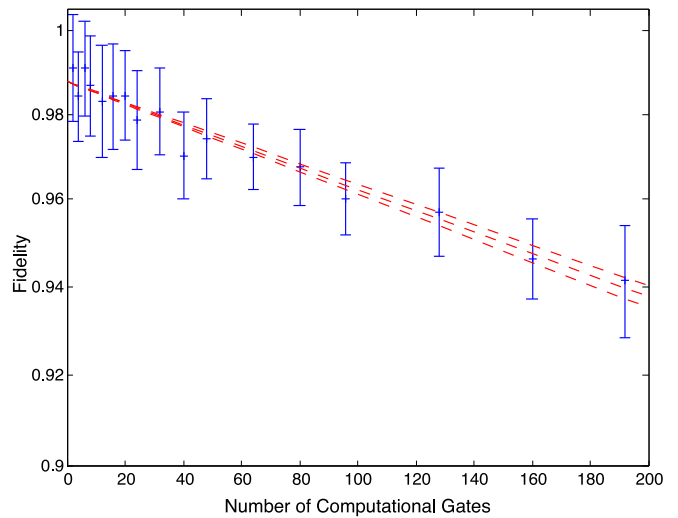}
\caption{\label{singlequbit_rf}
(Color online) Semi-log plot of the average fidelity as a function of the number of randomized gates for a single qubit using BB1 composite pulses after a r.f. selection sequence.  The error bars (68\% confidence) indicate the uncertainty from randomization (i.e. different computational sequences and Pauli randomizations give different fidelities due to coherent or biased errors).  The uncertainty in each measurement due to signal to noise and fluctuations in the amount of signal from the r.f. selection sequence is less than 0.5\%.   The fidelity decay is a good fit to a single exponential shown in red (dashed line) with 68\% confidence fits and reveals an error per gate of  $1.3 \pm 0.1 \times 10^{-4}$.}
\end{figure}

In NMR, the issues arising from r.f. inhomogeneity can be largely eliminated by running a r.f. selection sequence.  This is a sequence of pulses and gradients that leaves polarization on only a subset of the ensemble of processors that experience an r.f. field within a certain range, say $\pm 2\%$ of the ideal field strength \cite{Knill:2000a}.  For calibration purposes and again to avoid the sharp transitions of hard pulses we developed a numerically optimized control pulse which implemented the r.f. selection.  The pulse was designed to rotate spins outside the $\pm 2\%$ range of desired powers to the $x-y$ plane while leaving the calibrated spins along the $z$-axis.  The unwanted spins were then dephased using magnetic field gradient techniques.  This dramatically improves the results and gives a single exponential decay which we fit to give an error per randomized computation gate of $1.3 \pm 0.1 \times 10^{-4}$ (see Figure \ref{singlequbit_rf}).  A drawback of the r.f. selection sequence is that small fluctuations in the pulse power from the amplifier or changes in the resonant circuit give large (up to 5\%) changes in the output signal.  These were normalized through a stroboscopic observation of the signal after r.f. selection for each experiment.

An estimate of the expected error rate due to intrinsic decoherence can be made from the measured $T_1$ and $T_2$ values.  The combined time for a randomized computational gate using BB1 composite pulses is $516.8\mu s$ (including delays between pulses to avoid overheating).  A map consisting of purely $T_1$ and $T_2$ decoherence acting for this time would imply an error per randomized gate of $5 \times 10^{-5}$.   This represents a lower bound on the expected error rate which we should be able to reach with hardware and software improvements.  If the $T_2^*$ rather than the $T_2$ is used in the decoherence model, the estimated error per gate climbs to $4 \times 10^{-4}$.    The randomized gate sequence will somewhat refocus the static field inhomogenities contributing to $T_2^*$, but they are not explicitly refocussed.  The remaining impediments of incoherence across the ensemble members and the fluctuations in power from the amplifier could be overcome with even more robust and compensated pulses, although there is a tradeoff between more highly compensated pulses and the increased losses due to instrinsic decoherence because of the longer pulse times.  

For comparison purposes, we also tested other pulse types with the same protocol.  Using only simple uncompensated gaussian pulses we obtain an error rate of $2.1 \pm 0.1 \times 10^{-4}$ and using GRAPE  numerically optimized pulses \cite{Khaneja:2005fk}, an error rate of $1.8 \pm 0.2 \times 10^{-4}$.  The GRAPE pulses were numerically optimized to 99.999\% fidelity (Hilbert-Schmidt (HS) norm) over a range of r.f. powers $\pm 3\%$ from the ideal power.  They were $100\mu s$ in length and discretized at $1\mu s$.    It is somewhat  surprising that the numerically optimized pulses cannot match the performance of the BB1 pulses.  However, the BB1 pulses are well suited to compensating for systematic deviations from the ideal pulse shape which manifest themselves as calibration errors.  Numerically optimized pulses are somewhat robust to noise in the pulse generation:  because the controls are at a local maximum of fidelity, any deviation gives no change in the fidelity to first order.  However, numerically optimized pulses are still more sensitive to other imperfections in the implementation.  For example, the optimization and robustness assumes the control fields are constant at each time step in the discritized pulse.  In the experiment, finite bandwidth effects and noise prevent exact implementation of this and lead to a loss of fidelity.

\subsection{Multiple Qubits}
\begin{figure}[htb]
\includegraphics[scale = 0.65]{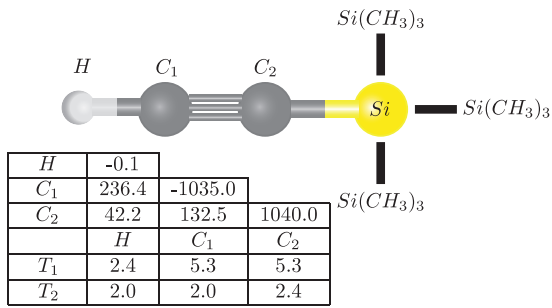}
\caption{\label{TMMS_Hnat}
(Color online) Structure of tris(trimethylsilyl)silane-acetylene and a table of natural Hamiltonian parameters (Hz) obtained from spectral fitting.  The diagonal elements give the chemical shifts with respect to the transmitter frequencies while the off-diagonal elements give the J-couplings.  $T_1$'s and $T_2$'s (seconds) are measured from standard inversion recovery and CPMG echo sequences respectively.  $C_1$ and $C_2$ were isotopically labelled with $^{13}C$ and rest of the molecule contained natural abundances and was ignored for the purposes of this experiment.}
\end{figure}

A three qubit molecule was made from a sample of selectively labelled $^{13}$C  tris(trimethylsilyl)silane-acetylene dissolved in deuterated chloroform \cite{Henry:2007a}.  The structure and a table of the natural Hamiltonian parameters is shown in Fig. \ref{TMMS_Hnat}.

%\begin{table}[htbp]
%\caption{\label{TMMS_Hnat} Table of natural Hamiltonian parameters (Hz) obtained from spectral fitting.  The diagonal elements give the chemical shifts with respect to the transmitter frequencies while the off-diagonal elements give the J-couplings.  $T_1$'s and $T_2$'s (seconds) are measured from standard inversion recovery and CPMG echo sequences respectively. } 
%\begin{ruledtabular}
%\begin{tabular}{|c | c | c | c|}
%\hline
% 		& 	$H$	&	$C_1$	&	$C_2$	\\
%\hline
%$H$		&	-0.1	&			&			\\
%\hline
%$C_1$	&	236.4&	-1035.0	&			\\
%\hline
%$C_2$	&	42.2	&	132.5	&	1040.5	\\
%\hline
%\hline
%$T_1$	&	2.4	&	5.3		&	5.3		\\
%\hline
%$T_2$	&	2.0	&	2.0		&	2.4		\\		
%\hline
%\end{tabular}
%\end{ruledtabular}

%\end{table}

Control was achieved through the GRAPE optimal control technique \cite{Khaneja:2005fk}.  The pulses were optimized to above 99.95\% HS fidelity over a range of r.f. powers $\pm 3\%$ from the ideal power.  The pulses were discretized at 2 $\mu s$ as a balance between smoothness  and spectrometer memory constraints. Single qubit pulses were $1.2ms$ long; CNOT gates between $H$ and $C_1$ (with any single qubit gate on $C_2$) were $2.4ms$; and CNOT gates between $C_1$ and $C_2$ (with any single qubit gate on $H$) were $4ms$.  These pulses  are not time-optimal but have low enough powers for experimental implementation.  Shorter pulses tended to require unfeasible high power levels which lead to probe heating during long computational sequences.  Non-linearities in the pulse generation and transient effects from the probe's resonant circuit lead to distortions in the implementation of shaped pulses.   To avoid this, the r.f. field at the sample was detected through a pickup coil and corrected through a simple feedback loop.  This correction procedure was only applied to individual pulses and the longer term power inverse droop  we observed  \footnote{The r.f. power from our signal generation and amplifier combination increases several percent over a timescale of 10's of $ms$.  See Ref. \cite{Skinner:2004th} for another observation of this effect in the context of optimal control.} was not corrected but should instead be handled by engineering robust pulses.   Due to finite spectrometer memory we were limited to 120 gates in a computational sequence.  Each truncation was averaged over 48 different computational gate sequences.   The same numerically optimized r.f. selection sequence used in the single qubit experiment was applied before each experiment to the proton nuclei.  Polarization on the carbon nuclei was dephased with gradient techniques giving the starting deviation density matrix $ZII$ (using product operator notation).  

A sequence of random gates was constructed in the following manner.  The Clifford group generating set was chosen to be the Hadamard and  $PHP^\dagger$ (a Hadamard gate conjugated by a phase gate) single qubit gates and CNOT gates between nearest neighbors.  With a probability of 2/3, a random single qubit gate was performed and with probability 1/3, a random CNOT was implemented  \cite{PhysRevA.70.052328}.  The resulting state was then tracked and a recovery sequence to return the state to $\pm ZII$ calculated. To design the recovery sequence, Hadamard or PHP$^\dagger$ gates were applied to each qubit such that their individual state was either $I$ or  $Z$.  This state was then transformed into the final $ZII$ by finding minimal amount of CNOT gates needed to transfer all the polarization back to the first qubit.  The algorithm is general and efficient in the number of qubits. These final recovery gates were not counted in the total number gates and will not affect the asymptotic error rate.  The entire sequence was then parallelized with a simplistic interative scheme of repeatedly checking whether gates in series could be compressed into a single parallel gate.  For example, a CNOT gate between qubits 2 and 3 followed by a Hadamard gate on qubit 1 would be compressed to a single timestep which implements both gates in parallel.  The fidelity of the state was then measured through a readout pulse on the proton spin.

\begin{figure}[htb]
\includegraphics[scale = 0.35]{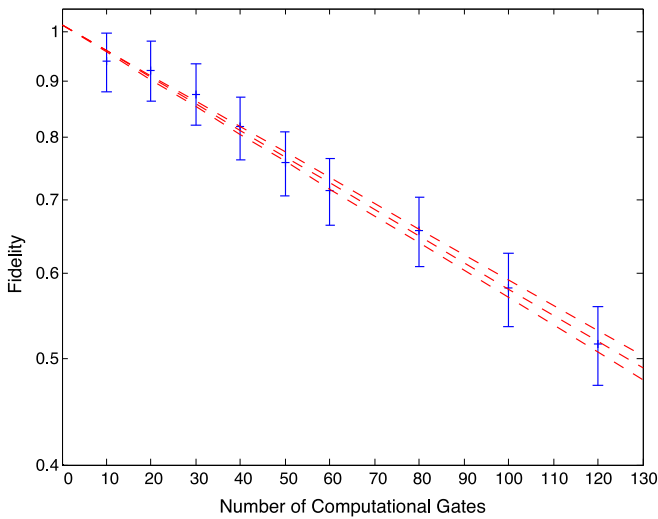}
\caption{\label{multiqubit}
(Color online) Semi-log plot of the average fidelity as a function of the number of randomized gates for the 3 qubit benchmarking experiment.  The error bars are obtained from the statistical nature of the randomization and the fitting of the NMR spectra.  To within the error bars a single exponential provides a good fit and gives an error per gate of $4.7 \pm 0.3 \times 10^{-3}$.}
\end{figure}

The results are shown in Figure \ref{multiqubit}.  The results fit an exponential decay well and give an error per gate of $4.7 \pm 0.3 \times 10^{-3}$, approximately an order of magnitude larger than the single qubit results. Again, an estimate of the lower bound on the error rate can be obtained from the measured $T_1$'s and $T_2$'s.  Assuming an independent and uncorrelated error model (which is unlikely but does not significantly affect the result) gives an average error per gate of $1.5 \times 10^{-3}$.  Moreover, from the design of the pulses, we would expect an error of $4.4 \times 10^{-4}$, which is an order of magnitude smaller than the experimentally measured error rate.  This leads us to suspect that there are still errors in the implementation of the pulses and/or knowledge of the chemical properties of the molecule that are not currently handled by our pulse design.\\

\section{Extensions to the Multi-qubit protocols}
\label{multiqubit_extensions}

More detailed information about the errors can be obtained by combining the ideas of previous randomization protocols \cite{Emerson:2007lr,levi:022314} with the randomized computational sequences.  For example, one may wish to determine on which qubits the errors are occurring or the difference in error rate between one and two qubit gates.  The steps of the proposed protocol are as follows:

\begin{enumerate}
\item Perform the single qubit benchmarking procedure on each qubit individually.  These numbers will give an estimate of the error per gate for single qubit gates.  Unfortunately, as discussed above, the error model is unlikely to follow an independent error model and the possibility that performing the single qubit gates induces errors on non-target gates needs to be checked.  This can be achieved by measuring the fidelity of the identity operation on the other $n-1$ qubits.  Efficient procedures exist for this measurement.  For example, performing single qubit Clifford gates at the beginning and their inverses at the end of the sequence and then randomizing allows an estimation of the fidelity of the channel with a small number of experiments \cite{Emerson:2007lr}.   A possible concern is that the error model on either the single qubit or the remaining $n-1$ qubits might be highly non-Markovian.  However, the benchmarking procedure should effectively act as a randomized dynamical decoupling sequence preventing entanglement between the two sub-systems \cite{viola:060502}.   

\item Two qubit benchmarking, using the procedure described above can then be performed on all pairs of qubits.  Knowing the single qubit error rates from the first step it should be possible to extract an estimate of the two-qubit error rate. The action on the other $n-2$ qubits should be characterized as in the first step to asses the fidelity of the wait steps.  

\item This procedure can be iterated to all groups of 3 qubits and so on but because most fault-tolerant constructions are specified in terms of one and two qubits gates, going as far as all pairs should be sufficient.

\end{enumerate}

\section{Conclusions}

We have demonstrated implementations of single and multi-qubit benchmarking in liquid state NMR.  In both instances the control is still not decoherence limited and improvements through both hardware and software should be possible. Potential software improvements include pulse more robust to calibration errors and noise in the pulse generation and better modeling of the system and apparatus.  Efforts in hardware improvements will be focussed on ensuring the implementation of the optimal control pulses is as close as possible to the ideal optimal control pulse.  

Simulations and proofs of fault-tolerant constructions suggest that given certain architecture assumptions, an error rate of $10^{-4}$ is sufficiently low to enable arbitrarily long quantum computations.  Here, the single qubit experiments demonstrated close to that level of control.  However, showing fault-tolerant levels of control on small demonstration systems does not imply a scalable quantum computer is possible.  Indeed the benchmarking of the three qubit system yielded an error per gate an order of magnitude worse than the single qubit system.  It is important to investigate how the level of control scales with the system size and how compatible the architecture is with the assumptions of the fault-tolerant construction before concluding anything about the fault-tolerance capabilities of a given system.  Multi-qubit benchmarking protocols will be an important part of that investigation.

\acknowledgements{
\emph{Author contributions - } C.R. ran the single qubit experiments.  M.L. analysed the r.f. inhomogeneity model.  M.L. and C.R. ran the multi-qubit experiments.  R.L., M.L and C.R. discussed the results and protocols.  C.R. and M.L. co-wrote the paper. 

The authors also wish to thank M. Ditty for technical assistance with the spectrometer and sample choice and David Cory for providing the 3 qubit sample.  We also thank J. Gambetta for useful questions and discussions and M. Silva, J. Emerson and E. Knill for stimulating discussions.}

\appendix

\section{Analysis of the r.f. inhomogeneity model.}
\label{rfanalysis_appendix}
As discussed in the experimental section, the main source of error in our liquid state NMR experiments is the r.f. field inhomogeneity across the sample.  This error model is not specific to NMR and can be applied to other systems: if a single instance of an experiment must be implemented multiple times, say, to reduce shot noise or measure the  expected value of a given observable, a parameter of the system can be constant for a single run of the experiment but fluctuate from one run to another.  In such a case, the final measurement will be related to an average over a distribution of that parameter. Here we show analytically how this error model gives a non-exponential decay in the single qubit experiment.

The consequence of the r.f. distribution is that not all the nuclear spins in the sample will have the same effective rotation under a nutation field.  For example, if we apply a r.f. field calibrated to rotate the spins by an angle $\frac{\pi}{2}$ about the $u$-axis ($u\in\{x,y,z\}$), the density matrix describing the state averaged across the sample in initial state $\rho$ is:
\be
\rho \to \int d\epsilon g(\epsilon)\Lambda_{P_u}(\epsilon)\left(e^{-i\frac{\pi}{4}P_u}\rho e^{i\frac{\pi}{4}P_u}\right),
\ee
where $\Lambda_{P_u}(\epsilon)$ is the superoperator describing the error for the spins experiencing a field $\epsilon$ away from the ideal field ( $\Lambda_{P_u}(\epsilon)(\rho)=e^{-i\epsilon\frac{\pi}{4}P_u}\rho e^{i\epsilon\frac{\pi}{4}P_u}$),  $g(\epsilon)$ is the r.f. distribution  and $P_u$ the appropriate rotation matrix.   The error model arising from r.f. field inhomogeneity is an over or under-rotation (by an amount $\epsilon$) along the same axis as the actual rotation.  With this notation the superoperator  describing a single instance of our single qubit experiment is  written as:
\begin{widetext}
\be
\label{actualchannel}
\begin{split}
\hat{\Lambda}_i(n) &= \int d\epsilon g(\epsilon) \hat{\Lambda}_{S_{i_n}}(\epsilon)\Sa_{i_n}\hat{\Lambda}_{P_{i_n}}
(\epsilon)\Pa_{i_n}\ldots\hat{\Lambda}_{S_{i_2}}(\epsilon)\Sa_{i_2}\hat{\Lambda}_{P_{i_2}}(\epsilon)\Pa_{i_2}\hat{\Lambda}_{S_{i_1}}(\epsilon)\Sa_{i_1}\hat{\Lambda}_{P_{i_1}}(\epsilon)\Pa_{i_1}  \\
&=\int d\epsilon g(\epsilon) \hat{\Lambda}_{i_n}(\epsilon)\Sa_{i_n}\Pa_{i_n}\ldots\hat{\Lambda}_{i_2}(\epsilon)\Sa_{i_2}\Pa_{i_2}\hat{\Lambda}_{i_1}(\epsilon)\Sa_{i_1}\Pa_{i_1},
\end{split}
\ee
\end{widetext}

where ${\Lambda}_{S_{i_j}}(\epsilon)(\rho)=e^{-i\epsilon\frac{\pi}{4}Q_{i_j}}\rho e^{i\epsilon\frac{\pi}{4}Q_{i_j}}$, ${\Lambda}_{P_{i_j}}(\epsilon)(\rho)=e^{-i\epsilon\frac{\pi}{2} P_{i_j}}\rho e^{i\epsilon\frac{\pi}{2} P_{i_j}}$.  $\hat{\Lambda}_{i_j}(\epsilon)$ is the cumulative error superoperator due to sequentially applying faulty $P_{i_j}$ and $S_{i_j}$.  Since the r.f. inhomogeneity error model is completely correlated with the pulses, the argument in Eq. \ref{emersonderive} cannot be directly applied. Nevertheless, Emerson et. al.  \cite{Emerson:2005a} conjectured that in the case of pulse dependent error, the cumulative noise operator after a sufficiently long sequence will become concentrated about some average value  (which we numerically verified in certain situations).  In the present case, the strength can be parametrized by the tipping angle of $\Lambda_{i_j}(\epsilon)$ and it can be easily verified that there are three relevant strengths, depending on whether $P_{i_j}$ and $S_{i_j}$ are along parallel, anti-parallel or perpendicular axes, as enumerated in Table \ref{strengthtable}.

\begin{table}[b]
\begin{tabular}{||c|c|c||}
\hline
\hline
Strength 		&	Axis			&	Probability	\\
\hline
\hline

$\frac{3\pi}{2}\epsilon$	&	Parallel		&	$1/8$\\
\hline
$\frac{\pi}{2}\epsilon$	&	Anti-parallel	&	$3/8$ \\
\hline
$\Gamma=2\cos^{-1}[\cos{(\frac{\pi}{4}\epsilon)}\cos{(\frac{\pi}{2}}\epsilon)]$	&	Perpendicular	&	$1/2$ \\	
\hline
\hline
\end{tabular}
\caption{\label{strengthtable} Table giving the three possible strength parameters for the cumulative error of a $\pi$ pulse followed by $\pi/2$ pulse due to r.f. inhomogeneity. Since the pulses are random and drawn from the set of 48 pulses described in Eq. \ref{PauliSimplectic}, each strength have a different probability of occuring.}
\end{table}

In term of depolarizing action, Eq. \ref{actualchannel} is equivalent to 
\begin{widetext}
\be
\label{effectivechannel}
\begin{split}
\hat{\Lambda}_i(n) &=  \int d\epsilon g(\epsilon) \Pa_{i_n}^\dagger\Sa_{i_n}^\dagger\hat{\Lambda}_{i_n}(\epsilon)\Sa_{i_n}\Pa_{i_n}\ldots\Pa_{i_2}^\dagger\Sa_{i_2}^\dagger\hat{\Lambda}_{i_2}(\epsilon)\Sa_{i_2}\Pa_{i_2}\Pa_{i_1}^\dagger\Sa_{i_1}^\dagger\hat{\Lambda}_{i_1}(\epsilon)\Sa_{i_1}\Pa_{i_1} \\
&=\int d\epsilon g(\epsilon)\prod_j\Pa_{i_j}^\dagger\Sa_{i_j}^\dagger\hat{\Lambda}_{i_j}(\epsilon)\Sa_{i_j}\Pa_{i_j}
\end{split}
\ee
\end{widetext}

The key observation to make is that each subset of $S$ and $P$ yielding a given noise strength parameter is sufficient to depolarize that given noise, e.g.
\be
\frac{1}{\mathcal{I}_{3\epsilon}}\sum_{\Sa,\Pa\in \mathcal{I}_{\frac{3\pi}{2}\epsilon}}\Pa^\dagger\Sa^\dagger\hat{\Lambda}_{\frac{3\pi}{2}\epsilon}\Sa\Pa =\hat{\Lambda}_{ave,\frac{3\pi}{2}\epsilon}.
\ee
where $\mathcal{I}_{\frac{3\pi}{2}\epsilon}=\{SP\,\, |\,\, S \textrm{ and }P \textrm{ are pulses along parallel axis}\}$ and ${\Lambda}_{ave,\frac{3\pi}{2}\epsilon}$ is the depolarized channel associated with the cummulative noise of strength $\frac{3\pi}{2}\epsilon$ with depolarizing parameter 
\be
p_{\frac{3\pi}{2}\epsilon}=\frac{4\cos^2{(\frac{3\pi}{4}\epsilon)} - 1}{3}.
\ee

Once the randomization over different gate sequences is performed, each $\Lambda_{i_j}$ in Eq. \ref{effectivechannel} will be randomized to a channel given by a weighted sum of the three different depolarizing channel, i.e.
\begin{widetext}
\be\label{averagedchannel}
\begin{split}
\hat{\Lambda}_{ave}(n) =& \frac{1}{|\mathcal{P}|^n|\mathcal{S}|^n} \sum_{i}\int d\epsilon g(\epsilon)\prod_{j=1}^n\Pa_{i_j}^\dagger\Sa_{i_j}^\dagger\hat{\Lambda}_{i_j}(\epsilon)\Sa_{i_j}\Pa_{i_j} \\
=& \int d\epsilon g(\epsilon)\prod_{j=1}^n\frac{1}{|\mathcal{P}||\mathcal{S}|} \sum_{i_j}\Pa_{i_j}^\dagger\Sa_{i_j}^\dagger\hat{\Lambda}_{i_j}(\epsilon)\Sa_{i_j}\Pa_{i_j} \\
=&\int d\epsilon g(\epsilon)\left[ \frac{1}{|\mathcal{I}_{\frac{\pi}{2}\epsilon|}} \sum_{\Pa,\Sa\in\mathcal{I}_{\frac{\pi}{2}\epsilon}}\Pa^\dagger\Sa^\dagger\hat{\Lambda}_\epsilon\Sa\Pa +\frac{1}{|\mathcal{I}_{\frac{3\pi}{2}\epsilon}|} \sum_{\Pa,\Sa\in\mathcal{I}_{\frac{3\pi}{2}\epsilon}}\Pa^\dagger\Sa^\dagger\hat{\Lambda}_{\frac{3\pi}{2}\epsilon}\Sa\Pa \right. +\left.\frac{1}{|\mathcal{I}_\Gamma|} \sum_{\Pa,\Sa\in\mathcal{I}_\Gamma}\Pa^\dagger\Sa^\dagger\hat{\Lambda}_\Gamma\Sa\Pa \right]^n  \\
=&\int d\epsilon g(\epsilon)\left[\frac{3}{8}\hat{\Lambda}_{ave,\frac{\pi}{2}\epsilon}+\frac{1}{8}\hat{\Lambda}_{ave,\frac{3\pi}{2}\epsilon}+\frac{1} {2}\hat{\Lambda}_{ave,\Gamma} \right]^n\\
=&\int d\epsilon g(\epsilon)\hat{\Lambda}_{ave}^n.
\end{split}\ee
\end{widetext}
Therefore, the effective averaged channel  action is given by
\be
\begin{split}
\hat{\Lambda}_{ave}(\rho) &=\bar{p}\rho +(1-\bar{p})\frac{\openone}{2}  \\
&	\bar{p}=\frac{3}{8}p_{\frac{\pi}{2}\epsilon}+\frac{1}{8}p_{\frac{3\pi}{2}\epsilon}+\frac{1}{2}p_\Gamma,
\end{split}
\ee

\begin{figure}[b]
\includegraphics[scale=0.35]{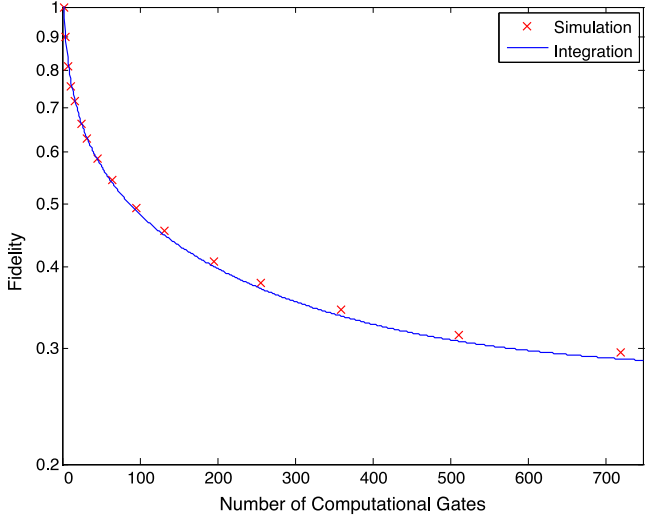}
\caption{\label{analyticaldecay}
Numerical simulation and analytical prediction of the fidelity decay of the randomized benchmarking protocol under a r.f. inhomogeneity error model plotted on a semi-log plot. The agreement of the two curves demonstrate the error model is well understood.  The small discrepancy is due to the finite number of runs in the numerical simulations.  This curve decays faster than that in Figure \protect{\ref{singlequbit_norf}} because this analysis uses simple pulses whereas the experiment used robust composite pulses.}
\end{figure}

The gate fidelity obtained by numerically integrating Eq. \ref{averagedchannel} using the measured r.f. distribution is compared to numerical simulations of the experimental sequences under the measured r.f. distribution in Fig. \ref{analyticaldecay}, which clearly demonstrate the non-exponential behavior of the decay.

\end{document}